# New Polynomial Law of Hadron Mass

**Sergio Bottini**

*Consiglio Nazionale delle Ricerche, Istituto di Elaborazione della Informazione,*

*I-56100 Pisa, Italy*

Abstract

A simple and general law of mass, intrinsically with zero freely-adaptable parameters, is shown to be valid for all the hadrons with one or more flavored ($s$, $c$, or $b$) quarks/antiquarks, both baryons and mesons. It establishes correspondences $H \Rightarrow \{P_i\}$ between these hadrons with at least one flavored constituent, $H$, and specific sets of lighter particles, $\{P_i\}$, in which the total baryon and lepton numbers are conserved. The law is polynomial with a varying degree that univocally depends only on the particles involved. Its statement may be given straightforwardly in terms of the binomial series converging to $(1 - \beta)^{-\frac{1}{2}}$. It asserts that, for each hadron $H$, there exists a certain set of lighter particles $\{P_i\}$, together with a specific reference hadron $h_0$ that fixes the mass scale, such that, for a value of the sum of this power series equal to the mass of $H$, a definite partial sum of the series equals the total mass of particles $P_i$. The starting, independent data in the rule are the masses of the hadrons with exclusively unflavored ($u$ and $d$) constituents (among which, our hadrons $h_0$, such as the proton or pion) and the masses of leptons. The consequence of the law is a pre-discretization of the mass spectrum of the hadrons containing at least one flavored constituent, on account of which the mass of any of them may only assume one of the values aprioristically determined by the total masses of lighter particles. By its simplicity, great accuracy, complete extensiveness and exhaustiveness, this rule may be regarded as a potential new physical law of mass.

## I. INTRODUCTION

Despite the increasing complexity of phenomena emerging in the various areas of science at deeper levels, nature sometimes appears to obey unexpectedly simple and elegant laws. Generally, a new law of nature seems to follow no logical necessity. Its statement may be completely unforeseen: initially guessed, it is then checked by the level of agreement with experimental data. Usually, only the pragmatic assumption that nature *can* be simple is the leading motivation behind any search for such laws.

On this basis, we have found that the masses of hadrons obey a simple relationship [1, 2, 3]. Our rule differs from a typical mass formula in that it consists of a condition on the masses which is only necessary. Thus, it appears to describe an *unsuspected* new property of hadron mass, fortuitously representable in the form of a law. The law is polynomial, essentially with no freely-adaptable parameter: only observed masses are involved. It concerns all hadrons, which are partitioned into two different roles depending on their quark contents: flavored or unflavored. According to this law, the masses of the hadrons with one or more flavored constituents, namely strange (*s*), charm (*c*), or bottom (*b*) quarks, and/or their respective antiquarks, may only take on their values within a certain set of permitted values, which are fixed by the masses of lighter particles. Such a pre-discretization of mass then represents the way in which our rule would take part in the generation of the mass spectrum of these hadrons. The masses of the remaining hadrons, i.e. those with exclusively unflavored constituents, namely up (*u*) and down (*d*) quarks/antiquarks, such as the proton and pion, are, together with the masses of leptons, the independent values in the rule, i.e. the starting data. The wide and complete validity of the rule, for both mesons and baryons, suggests that it may originate from a physical cause.

## II. NEW LAW OF MASS

### A. Definition

For each observed hadron (baryon or meson) containing one or more flavored ($I = 0$) quarks/antiquarks (namely, $s$, $c$ or $b$), indicated as hadron $H$, *there exists* a certain set of lighter particles, $\{P_i\}$, together with a specific hadron containing only unflavored ($I = \frac{1}{2}$) constituents (namely, $u$ and $d$) which fixes the mass scale, $\mathcal{M}$, called reference hadron $h_0$, i.e. *there exists* the correspondence

$$H \stackrel{h_0}{\Rightarrow} \{P_i\}, \tag{1}$$

*such that* the value $\beta = \beta_H$ for which

$$\mathcal{M}(H) = S(\beta_H) \tag{2}$$

also satisfies

$$S_{n+1}(\beta_H) = \sum_i \mathcal{M}(P_i), \tag{3}$$

where

$$S(\beta) = \frac{1}{\sqrt{1-\beta}} = \sum_{k=0}^{\infty} \binom{-\frac{1}{2}}{k} (-\beta)^k \tag{4}$$

(defined for $-1 \leq \beta < 1$); $S_{n+1}$ is a definite partial sum of this binomial series, i.e. the sum of its first $n + 1$ terms, with $n$ univocally depending only on the particles involved; and

$$\mathcal{M}(\cdot) = \frac{m(\cdot)}{m(h_0)}, \tag{5}$$

i.e. the masses $m(\cdot)$ of $H$ and $P_i$ are re-expressed adimensionally as $\mathcal{M}(\cdot)$ in units of the mass of $h_0$. This $h_0$ is a baryon or meson according to $H$, with $S = C = B = 0$, $I \neq 0$, and $L(h_0) = L(H)$.

In Eq. (1), the total baryon and lepton numbers and the total electric charge, between $H$ and $\{P_i\}$, are conserved. The types considered of these correspondences (which include cases formally analogous to those produced by hadronic, radiative or semileptonic decays with the minimum number of final hadrons) are

$$H \Rightarrow \begin{cases} P_1 P_2 & \text{generally, where the } P_i \text{ are lighter hadrons with any kind of quarks} \\ P_1(\gamma) & \text{only for the same quark content of } H \text{ and } P_1 \\ P_0 \mu(\nu/e) & \text{only if } P_0 \text{ is an } S = C = B = 0, I \neq 0 \text{ ground-state hadron} \end{cases} \quad (6)$$

with $L(P) \leq L(H)$ (note that $P_0$ contains only unflavored constituents). The value of $n$ for the $(n+1)$th partial sum of our power series, in Eq. (3), is determined by $H$, $P_i$ and $h_0$ through

$$n = \begin{cases} n_{min} + 1 & \text{if } H \text{ contains two or more flavored } (I = 0) \text{ quarks/antiquarks} \\ & \text{with the same absolute electric charge} \\ n_{min} & \text{otherwise,} \end{cases} \quad (7)$$

where $n_{min}$ is the minimum integer for which

$$S_{n_{min}+1}(1 - \varepsilon) > \max\{\sum_i \mathcal{M}(P_i)\}, \quad (8)$$

with $(\beta =)\ 1 - \varepsilon$ being a test-value very close to one (where these partial sums approach their upper bounds), let be $\varepsilon = a \cdot \exp[-b\ (n_{min} + 1) \mathcal{M}^2(H)]$ with $a = 2.97 \cdot 10^{-3}$ and $b = 3 \cdot 10^{-5}$ (note that for $\beta_H \lesssim 0.99$ it will be $\varepsilon \sim a$ for all $n_{min}$ of interest).

The precise values to be used for the masses of the hadrons $H$ and $P_i$ in Eqs. (2) and (3) are the average values of the masses of those isospin states of them that intervene in the various charge-conserving variants of the same instance of correspondence (1). On the right-hand side of Eq. (8), we consistently have the largest value of the total mass of each $\{P_i\}$ (depending on the permitted isospin states of the hadrons involved). We specify that for the value of $m(h_0)$ in Eq. (5) in the cases of the nucleon and pion we take the mass of the proton ($p$) and the neutral pion ($\pi^0$), i.e. the lowest-lying baryon and meson, respectively. Finally, for the hadrons $H$ with excitations of a radial type, the relative quantum numbers $N$ follow: $N(h_0) \leq N(H)$ and $N(P) \leq N(H)$.

This law of mass can be summarized as:

*Statement:*

$$\forall\ H\ \exists\ \{P_i\},\ h_0\ |\ S_{n+1}(\beta_H) = \sum_i \mathcal{M}(P_i)\ \text{ for }\ \mathcal{M}(H) = S(\beta_H) = (1 - \beta_H)^{-\frac{1}{2}} \quad (9)$$

with $\mathcal{M}(\cdot) = m(\cdot)/m(h_0)$, where $\{P_i\}$ is found to satisfy Eq. (6), and $n$ to obey Eqs. (7) and (8), i.e. $n$ is univocally determinable *a priori*, independently of Eq. (3), as a function of the content of flavored quarks/antiquarks of $H$ and of the magnitude of the total mass of particles $P_i$ in units of $m(h_0)$.

### B. The rule re-written as a polynomial law

Our law of mass can be expressed equivalently in the form of a polynomial law of a varying degree, as follows: for each $H$ there exist $\{P_i\}$ and $h_0$ such that

$$Q_n[\mathcal{M}^{-2}(H)] = \sum_i \mathcal{M}(P_i) \qquad (10)$$

is satisfied, where

$$Q_n[\mathcal{M}^{-2}(H)] = \sum_{k=0}^{n} \binom{-\frac{1}{2}}{k} [\mathcal{M}^{-2}(H) - 1]^k \qquad (11)$$

is the polynomial function of degree $n$ in the dimensionless variable $\mathcal{M}^{-2}(H)$, with $0 < \mathcal{M}^{-2}(H) \leq 2$, obtained by substituting $\beta_H$ from Eq. (2), i.e.

$$\beta_H = 1 - \mathcal{M}^{-2}(H), \qquad (12)$$

into Eq. (3). Degree $n$ is fixed *a priori* by Eqs. (7) and (8), with $Q_{n_{\min}}(\varepsilon)$ replacing the left-hand side in Eq. (8). The mass scale $\mathcal{M}$ is set by $h_0$ through Eq. (5).

Conversely, property

$$\lim_{n \to \infty} Q_n[\mathcal{M}^{-2}(H)] = \mathcal{M}(H) \qquad (13)$$

of polynomials (11) makes it possible to re-transform this alternative formulation of the law, Eq. (10), into Eqs. (2) and (3).

### C. A more detailed criterion for *n*

Although it is not essential for the complete validity of the rule, we wish to take into account a minor variant of the criterion that establishes the value of $n$ in Eq. (7), exclusively for values of $\beta_H$ very close to one, i.e. $0.99 \lesssim \beta_H < 1$. Specifically, for $\beta_H > \hat{\beta}(n) = 1 - c \cdot \exp[-d \cdot (n+1)]$, with $c = 1.7 \cdot 10^{-2}$ and $d = 4.8 \cdot 10^{-2}$, the following

$$n = n_{min} \quad (\text{with } \beta_H \approx 1) \quad (14)$$

holds for any quark content of *H*, replacing the specialized Eq. (7). The few cases falling under this additional condition will be identified.

### D. Comments and specifications

In its definition, our rule has essentially zero freely-adaptable parameters. This is strictly true for the whole sector of known baryons. Note that $\beta_H$ is a simple representation of the experimental mass of *H*, according to Eq. (12). Only in the choice of the small $\varepsilon$ forming the test-value of $\beta = 1 - \varepsilon \ (\approx 1)$ in Eq. (8), is there a certain, though low, level of arbitrariness. However, this may generally have a just slight effect on the determination of *n* only for quite high values (i.e., for some mesons, when also $\beta_H \approx 1$).

In almost all cases, we will have $0 < \beta_H < 1$, and accordingly $m(H) > m(h_0)$. The permitted negative values of $\beta_H$ (i.e. $-1 \leq \beta_H < 0$) make it possible to cover the rare cases of the hadrons *H* which are lighter than any possible reference hadron $h_0$ with the same *L*, for $m(h_0)/\sqrt{2} \leq m(H) < m(h_0)$, thus preserving our condition $L(h_0) = L(H)$. This happens, for example, for the $L = 1$ baryon $\Lambda(1405)$ $(= H)$ which is lighter than the lowest-lying $L = 1$ baryon with only unflavored constituents, $N(1520)$ or $N(1535)$ $(= h_0)$. However, although it is interesting that these cases too have a solution in the rule still with $L(h_0) = L(H)$, their very low number may not rule out the possibility that condition $\beta_H > 0$ should always be satisfied. As a matter of fact, alternative solutions with $\beta_H > 0$, i.e. $m(H) > m(h_0)$, and by necessity here with $L(h_0) < L(H)$, would also exist for such few cases.

In defining Eq. (6), we referred in passing to our correspondences $H \Rightarrow \{P_i\}$ as including the formal analogues of decays of hadrons *H* only because this representation rapidly conveys some correct information about the kinds of correspondences permitted in the rule. Therefore, the involvement of the massless $\gamma$ as well as the neutrinos is purely nominal in this context. Our rule would be consistent with the limitation to the correspondences of type (6) which, apart from non-massive particles, may formally be obtained through one or more steps of possible decays. However, to be specific, only the conservation of the baryon number and electric charge is strictly imposed on Eq. (1). In this following, we will accept any correspondence (1) which satisfies these conditions, under the restriction that $\{P_i\}$ represents two terms, $P_1$ and $P_2$, of which at least one is always a hadron and the other, when it is not a hadron itself, may be, in accordance with the cases described by Eq. (6), either an essentially massless term or a term characterized by the mass of the muon. For example, given $\Sigma^0 \Rightarrow \Lambda\gamma$, we may consider $\Sigma^0 \Rightarrow \Lambda e^+ e^-$ as substantially equivalent; and hence $\Sigma^\pm \Rightarrow \Lambda e^\pm \nu$.

In the application of the criterion of assignment of $n$, Eq. (7), to the particular case of the mixed states ($u\bar{u}$, $d\bar{d}$, $s\bar{s}$), we may consistently expect $n = n_{min}$ or $n_{min} + 1$ depending on whether the percentage of the amount of $s\bar{s}$ content is less than or greater than a certain threshold in between fifty and a hundred per cent. We will assign $n = n_{min}$ to mesons $\eta(548)$ and $\eta'(958)$, as well as to $f_o(980)$, and $n = n_{min} + 1$ to the almost pure $s\bar{s}$ states, such as $\phi(1020)$ and $f_2'(1525)$. For certain $I = 0$ mesons, the percentage content of their strange constituents is a controversial matter. For the $\eta$ states with masses above 1 GeV, we will take $n = n_{min} + 1$.

Finally, note that the $(n+1)$th partial sums in Eq. (3), $S_{n+1}(\beta_H)$, are represented, in their respective domains implicitly given by Eqs. (7) and (8), by polynomial curves whose effective orders range from the first to, at most, the third, irrespective of the value, even if high, of $n$. In fact, when expressed in powers of $1-\beta_H$, they become

$$S_{n+1}(\beta_H) = \sum_{k'=0}^{n} (1-\beta_H)^{k'} \cdot \sum_{k=k'}^{n} (-1)^{k-k'} \cdot \binom{k}{k'} \cdot \binom{-\frac{1}{2}}{k}, \qquad (15)$$

where any terms of a higher degree than the third can always be neglected for our purposes (i.e., $k' \leq 3$) because, as is easy to see, the largest value that variable $1-\beta_H$ may assume sufficiently approaches zero with $n$ increasing. By substituting $1-\beta_H$ with $\mathcal{M}^{-2}(H)$, Eq. (15) gives the corresponding expression of polynomial $Q_n[\mathcal{M}^{-2}(H)]$ in powers of $\mathcal{M}^{-2}(H)$.

## III. AGREEMENT WITH THE DATA

All the known baryons and mesons with one or more flavored constituents, i.e. hadrons $H$, are found to satisfy our law, Eqs. (2) and (3), or equivalently the single Eq. (10), with a very good approximation. The respective sets of particles $\{P_i\}$ are listed in Table I for $L(H) = 0$ and, for some significant cases, in Table II for $L(H) > 0$. The rule is illustrated graphically in Fig. 1. A comprehensive view of its excellent agreement with the data is cumulatively given in Fig. 2 for all our reference hadrons $h_0$.

To show the high degree of accuracy of the rule, we have derived the mass values of hadrons $H$, $m_{calc}(H)$, from the experimental masses of the corresponding (lighter) particles $P_i$ (see Tables I and II). Small variations in the mass values of the $P_i$ correspond, in fact, to larger (and, with $\beta_H$ approaching 1, increasingly larger) changes in the calculated masses of the $H$'s, by a factor ($> 1$) equal to the ratio between the derivatives of the sum of series (4) and of its $(n + 1)th$ partial sum for $\beta = \beta_H$, i.e. $S'(\beta_H)/S'_{n+1}(\beta_H)$. Thus, in comparing the rule with the data, the sum of the masses of the particles of a given set $\{P_i\}$ is taken as an input, and, accordingly, the mass of $H$, calculated through Eq. (10), is the output of the rule. The experimental uncertainties on the masses which enter the calculations produce the uncertainties associated with the values of $m_{calc}(H)$. The experimental masses are taken from [4] (more recent data for the $\Lambda_b$ [5] and $\Sigma_c^*$ [6] are available). Recently, CLEO [7] reported the discovery of the ($J^P = \frac{3^+}{2}$) $\Xi_c^*$ whose mass further confirms the validity of our rule.

Because of the experimental uncertainties, it should be realized that a reference hadron $h_0$ whose mass is not known with great accuracy may introduce a small margin of choice. Clearly, in the important cases of the highly accurate reference masses of the proton and $\pi^0$ such an arbitrariness is reduced to zero. Moreover, for relatively low values of $\beta$, the exact determination of a reference mass inside its uncertainty interval is not crucial. To form $m(h_0)$ in Eq. (5), we might, in general, establish to take the $I = 1/2$ or $I = 0$ state of $h_0$ according to whether it is a baryon or meson. In the Tables, we have precisely used the experimental mean/nominal values of mass for all the reference hadron states considered except, to a very small (and justified) extent, in the case of $h_0 = \rho(770)$ (see Table I), for which, in order to show the best fit to data, we have expressed a certain, though restricted, choice within its experimental uncertainty. The optimal value turns out to be 766.64 MeV, that we assume to refer to the neutral state $\rho^0$ (with the mass of $\rho^\pm$ taken, say, 1.5 MeV higher). Consider, however, that just using the experimental mean values for the $\rho(770)$, too, would still very well support the evidence for the complete validity of the rule (note that the mean experimental data for $\rho^\pm$, $766.9 \pm 1.2$ MeV, is very close to our value). Therefore, the rule does not rely essentially on any adjustable parameter.

For the sake of completeness, consider that the $(I = 0)$ $\omega(782)$ might, in principle, be included as a possible additional $L = 0$ reference meson since, like $\rho(770)$, it contains only unflavored constituents. In particular, for $h_0 = \omega$, correspondences $K^* \Rightarrow \rho\mu(\nu)$ and $\phi \Rightarrow K\bar{K}$, found for $h_0 = \rho$ (Table I), would be even more accurately satisfied, with 894 and 1018 MeV calculated vs. 894 and 1019 MeV experimental for the masses of $K^*$ and $\phi$, respectively.

**IV. SIDE CONSIDERATIONS**

**A. Probability of fitting the data by chance alone**

The combinations of particles $P_i$ permitted by Eq. (6), when used as inputs in Eq. (10), yield solutions which include the mass spectrum of the hadrons $H$ as a subset. An estimation of how likely a full agreement would be between our rule and the data if it were to occur only by chance, to the same accuracy as our fit, for the well-established $L = 0$ ground-state hadrons $H$ alone, with $h_0 = \pi^0/\rho$ for mesons and $h_0 = p$ for baryons, gives a probability of less than $10^{-16}$. Although this probability is ideal, in that it has been calculated by replacing the outputs of the rule with as many random outcomes that have a uniform distribution, nevertheless it is clearly indicative of our case. Such an extremely small value for the probability of fitting the data accidentally in our rule indirectly supports the conjecture that it may have a physical nature.

## B. Counts of coincidences for fictitious reference masses

By a computer analysis, we directly checked that the masses of the proton, pion and $\rho(770)$, in their role as $L = 0$ reference hadrons $h_0$, are really special values for the rule, in that they produce sharp peaks in the number of the $L = 0$ hadrons $H$ that satisfy correspondences $H \Rightarrow \{P_i\}$ in Eq. (10), against the number of accidental coincidences obtained by taking any other, arbitrary value as a (fictitious) trial reference mass (see Fig. 3). For the mesons, the test was made when the available experimental data were those of [8], which however, for our purposes, are essentially equivalent to the present data of [4] (used for baryons). The hadrons taken as $P_i$ were ground-state hadrons.

Thus, the rule has the remarkable ability of finding out the masses of hadrons with only unflavored constituents, on the basis of their property (when taken as reference masses) of yielding good solutions in Eq. (10) for the hadrons $H$ concerned (as soon as a sufficiently high number of these hadrons H is known).

This is also confirmed in the case of $L = 1$, where the optimal agreement with the data is reached for a reference mass in the region of $N(1520)/N(1535)$ for baryons (namely 1526 MeV), and $b_1(1235)/a_1(1260)$ for mesons (namely 1228 MeV, still within the smaller experimental uncertainty of $b_1$), i.e., respectively, the lightest $L = 1$ baryon and meson with only unflavored constituents.

## C. Generative criterion of mass spectra of $L = 0$ ground-state hadrons $H$

Our law consists of a necessary, but not sufficient, condition for a given value to be the mass of a hadron $H$. Therefore, it cannot be seen as a typical mass formula which reproduces a certain spectrum by itself. However, in order to provide a qualitative evaluation of which part the rule appears to take in the formation of the mass spectrum of the hadrons with at least one flavored constituent, we have found that, for example, in the important case of the known ground-state $J^{PC} = 0^{-+}, 1^{--}$ mesons $H$ and $J^P = \frac{1^+}{2}$ baryons $H$, by adding a few fair constraints of an essentially topological character to this law of hadron mass, we can generate the masses correctly.

The generative criterion of these masses can be described through a process of growth of directed graphs, the rooted trees shown in Fig. 4, where roots, vertices, and edges represent particles. Specifically, as roots: the same hadrons that may be our reference hadrons ($h_0$); as vertices: hadrons $H$; and as edges: any particles $P_i$. These graphs develop upwards from their roots, producing vertices $H$, consistently with the requirement that correspondences $H \Rightarrow \{P_i\}$ are read downwards, i.e. in the inverse direction. Thus, for each vertex $H$, the respective set $\{P_i\}$ is constituted by the particles which represent the adjacent lower vertex/root and their joining edge. With each of these vertices $H$ we associate a weight, $w(H)$, defined as the total mass of the final particles, excluding the root, that are obtained by iterating the application of $H \Rightarrow \{P_i\}$, sequentially, also to those $P_i$ which still contain flavored constituents, if any, until only hadrons with exclusively unflavored constituents, or leptons, or $\gamma$'s remain. For example, from the cascade $\Xi \Rightarrow \Lambda\pi$, $\Lambda \Rightarrow N\pi$, we obtain $N\pi\pi$ as final particles, and then, excluding root $N$, $w(\Xi) = 2m(\pi)$. We see that the contributions to the weight assigned to a vertex $H$ come from the edges in the chain connecting it to the root.

The starting data are the masses of the $L = 0$ ground-state hadrons with only unflavored constituents (namely $\pi$, $\rho$, $\omega$, $N$ and $\Delta$), and the masses of the leptons (essentially, the $\mu$). The hadrons $H$ generated as new vertices may be re-used as $P_i$, i.e. lower vertices or edges, to keep on climbing the mass spectrum. The trees for the hadrons with the lesser $J$ (i.e. 0 for mesons and 1/2 for baryons) are formed first.

For the $L = 0$ hadrons $H$ considered, our correspondences $H \stackrel{h_0}{\Rightarrow} \{P_i\}$ are correctly selected by taking the next higher outputs of the rule under the following constraints:

*-i) for a vertex H being the lowest-lying meson/baryon with one or two constituents of the same given flavor: $\Delta w$ = min, with a number of ending consecutive edges of $\pi$-type at least equal to the number of its flavor unities;*

*-ii) J(H) = lesser J only if $\pi/N \in \{h_0, P_i\}$ for mesons/baryons H;* and finally (with rare occurrences)

*-iii) maximizing the outputs within small ranges: if two sets of particles, $\{P_i\}$ and $\{P_i'\}$, with the same flavors and values of $I_3$ involved, yield close outputs for the same $h_0$, say to within 100 MeV, the higher one is preferred;*

where $\Delta w$ is the increase in weight $w(H)$ relative to the case with the same number of quarks/antiquarks of the right lighter flavor, and *min* stands for a minimum value of mass equal to the total mass of one or more massive particles.

For example, for only one flavored constituent (i.e. for $K$, $D$, $B$, $\Lambda$, $\Lambda_c$, $\Lambda_b$), it is $\Delta w_s = \Delta w_b = m(\pi)$ and $\Delta w_c = m(\mu)$ for mesons, and $\Delta w_s = \Delta w_c = m(\pi)$, $\Delta w_b = m(\mu)$ for baryons. Still as an example, point *iii* implies that $K\eta'$ (with a slightly higher output) is preferred to $K^*\eta$ for $H = D^*$, with $h_0 = \rho$. It should be noted that the only constraints *i* and *ii* would suffice for all the $L = 0$ ground-state mesons and baryons with the lesser $J$ up to a content of two flavored constituents. The extension to the $J^P = \frac{3+}{2}$ baryons $H$ would require the addition of a few other general constraints (already satisfied spontaneously by all the above $L = 0$ hadrons $H$).

As supplementary cases, Fig. 4 also reports the selected correspondences for mesons $B_c$, $B_c^*$, and $\eta_b$, and baryon $\Xi_{cc}$, which predict plausible values for their masses (see Table III). The further predictions of the graphs would also be consistent with the expected values for the masses of the hadrons with top quarks, although, in all these cases, the outputs of the rule are, because of their very high values, very strongly dependent on the variations of the masses in input. For example, from $\Lambda_t \Rightarrow \Lambda_b \pi^+$, with $h_0 = p(938)$ and $n = 29$, we would have $m(\Lambda_t) = 181$ GeV for a value of the mass of $\Lambda_b$ in input of 5633.7 MeV.

This adjunctive criterion simply shows how a small set of essentially not quantitative constraints may be sufficient, in many significant cases, to correctly select the masses of the hadrons with at least one flavored constituent from the set of the values which are pre-determined by our law, Eq. (9), as their possible candidates. In other words, it is as though these constraints were a "measure" of how much information our law would need (by its nature of being an only necessary condition on the masses) in order to reproduce the mass spectrum of these hadrons $H$ properly.

**V. SUMMARY AND CONCLUDING REMARKS**

We have shown that it is possible to define a rule, essentially with zero freely-adaptable parameters, which maps all the observed hadrons with at least one flavored constituent, $H$, onto specific sets of lighter particles, i.e. $H \Rightarrow \{P_i\}$, Eq. (6), under the condition that the total baryon and lepton numbers and the total electric charge are conserved. With each of these correspondences, a specific reference hadron $h_0$ is associated containing only unflavored constituents, such as the proton or pion, which fixes the mass scale $\mathcal{M}$, Eq. (5), with $L(h_0) = L(H)$. The rule asserts that if the mass value of a hadron $H$ is set equal to the sum of the binomial series $S(\beta) = \sum_{k=0}^{\infty} \alpha_k \beta^k = (1-\beta)^{-\frac{1}{2}}$, then a definite partial sum of this power series must equal the total mass of certain lighter particles $P_i$; i.e., the rule consists of mapping $H \Rightarrow \{P_i\}$ established by $S_{n+1}(\beta_H) = \sum_i \mathcal{M}(P_i)$ for $\mathcal{M}(H) = S(\beta_H)$, Eqs. (3) and (2). Alternatively, by combining together these two equations, which separately involve the partial and total sums of the series, the rule can be expressed equivalently as a polynomial law of a varying degree $n$, between the inverse-square of the mass of $H$ and the total mass of the $P_i$, namely: $\forall H \exists \{P_i\}, h_0 \mid Q_n[\mathcal{M}^{-2}(H)] = \sum_i \mathcal{M}(P_i))$, Eq. (10), with $\mathcal{M}(\cdot) = m(\cdot)/m(h_0)$, Eq. (5); where the coefficients of polynomial function $Q_n$, Eq. (11), are determined as being the first $n+1$ coefficients of our binomial series. Degree $n$ (or, equivalently, the point of truncation of the series) is a function of the triple $(H, \{P_i\}, h_0)$: specifically, $n$ is jointly and univocally fixed by the content of flavored quarks/antiquarks of $H$, Eq. (7), and by the magnitude of the total mass of particles $P_i$ in units of $m(h_0)$, Eq. (8). We thus have a family of polynomial functions of which only one is selected *a priori* in order to satisfy each given correspondence $H \Rightarrow \{P_i\}$. Although $n$ can be large, our polynomials are represented in their domains by curves whose effective orders only reach, at most, the third one [see Eq. (15)].

    While the latter formulation of the law, given in terms of a single polynomial equation, is more concise and usual, the former formulation has the interesting property that all the masses are treated on the same footing, i.e. also the mass of $H$, Eq. (2), like

those of the $P_i$, Eq. (3), is included in a linear term. According to this form of the law, the hadrons with strange, charmed or bottom constituents appear to be endowed with an unsuspected new property by which specific fractions of their masses would be constrained to correspond in value to the total masses of certain lighter particles.

The rule expresses a necessary condition on mass, thus implying a prohibition: the values that do not satisfy this condition, Eq. (10), are forbidden as mass values for the hadrons with at least one flavored ($I = 0$) constituent. It has predictive capabilities: for example, as a partial recapitulation, the mass of meson $K$ is determined by (the mass of) the $\pi$; the mass of $\eta$ by the $\pi$ and $\mu$; the mass of $\eta'$ by the $\pi$ and $\rho$; the mass of baryon $\Lambda$ by the $N$ and $\pi$; the mass of $\Sigma$ by the $\Lambda$; the mass of $\Xi$ by the $\Lambda$ and $\pi$; the mass of $\Omega$ by the $\Lambda$ and $K$; the mass of $\Lambda_c$ by the $\Sigma$ and $\pi$; and so on. In general, for any mass interval expected to contain hadrons with one or more flavored constituents, the rule determines certain values which would be the only possible candidates for those masses. Thus, the whole mass spectrum of such hadrons would be "pre-discretized": their masses are a subset of the values permitted *a priori* by the rule on the basis of the masses of lighter particles. The law is in very good agreement with the experimental data, without exceptions. The high number of its accurate solutions can be regarded as a very sound guarantee against chance, i.e. in favor of the conjecture that it may actually reveal *physical* regularities of the hadron mass spectrum.

## Table Captions

**TABLE I**. Results of the fit to the masses of the $L = 0$ hadrons $H$, i.e. the $L = 0$ baryons and mesons with at least one flavored ($I = 0$) quark/antiquark, including the radially excited states. For each reported correspondence, $H \xRightarrow{h_0} \{P_i\}$, the calculated mass value of $H$, $m_{\text{calc}}(H)$, has been obtained from the experimental masses of the respective lighter particles $P_i$, through Eqs. (2) and (3), or, equivalently, the single Eq. (10). The value of $n$, which indicates the point of truncation of binomial series (4) in our equations, is determined by Eqs. (7) and (8). Hadron $h_0$ is the associated reference hadron which sets the mass scale in the rule. Also reported are the calculated values of the dummy parameter $\beta$, $\beta_{H,\text{calc}} = 1 - m^2(h_0)/m^2_{\text{calc}}(H)$, which only has a formal role in the law. As mass of $\eta(548)$ in input we used 547.93 MeV, taken within its experimental uncertainty range; the outputs from its current mean value, 547.45 MeV, are given in parentheses. A few hadrons $H$ are found to satisfy correspondences for both their possible reference hadrons $h_0$. [a] (Extra) cases which follow Eq. (14) for $n$.

**TABLE II**. Some examples of the fit to the masses of the hadrons $H$ with $L > 0$. The case of the controversial $L = 1$ meson $f_o(980)$ (= $H$) illustrates that imposing everywhere condition $\beta_H > 0$ would sometimes necessarily require $L(h_0) = L(H) - 1$; otherwise, $L(h_0) = L(H)$ can always be obeyed. [b] Underestimated value because we have taken zero for the unknown mass difference inside the isospin doublet of $N(1535)$.

**TABLE III**. Further values of mass generated, as predictions, by the adjunctive criterion forming the hadron trees in Fig. 4, through selection from the mass values permitted by our rule. [a] See footnote of Table I.

# Figure Captions

**FIG. 1**. Graphic representation of mapping $H \Rightarrow \{P_i\}$ established by the rule, in the case of the $L = 0$ ground-state baryons $H$ with $h_0 = p(938)$. The mass of $H$ is reported vertically on the left, the total mass of the corresponding particle set $\{P_i\}$ on the right. The scale of mass is set by the proton, being the current reference baryon ($h_0$), i.e. $\mathcal{M}(\cdot) = m(\cdot)/m(p)$. On the horizontal axis we have the values of $\beta_H$ which, as the left vertical axis, again represent the experimental masses of hadrons $H$, expressed in accordance with Eq. (12). The linear, quadratic, cubic, ... curves (labeled with $n = 1, 2, ...$) represent the consecutive partial sums of the binomial series with (total) sum $1/\sqrt{1-\beta}$ $(= \sum_{k=0}^{\infty} \alpha_k \beta^k)$, increasingly approaching their limit curve (in bold) as their degree $n$ increases (here shown up to $n = 6$). Our law of mass, expressed by Eqs. (2) and (3), or equivalently by Eq. (10), asserts that for every hadron $H$ there must exist one experimental point with coordinates $[\beta_H = 1 - \mathcal{M}^{-2}(H), \sum_i \mathcal{M}(P_i)]$ which belongs to the one of these curves that is jointly and univocally determined *a priori* by the flavored quark content of $H$ and the magnitude of the total mass of particles $P_i$, according to Eqs. (7) and (8) respectively.

**FIG. 2**. Cumulative view of the very good agreement of the rule with the data (reported in Tables I and II), shown for $n \leq 7$, in both sectors of baryons and mesons $H$, for all our reference hadrons $h_0$, in (a) for $n = n_{min}$ and in (b) for $n = n_{min} + 1$, depending on the flavored quark content of $H$, Eq. (7). Masses $\mathcal{M}$ are expressed in units of the mass of the relative $h_0$, and thus "normalized". For each $H \stackrel{h_0}{\Rightarrow} \{P_i\}$, only one polynomial curve of a certain degree $n$ $(= 1, 2, ...)$ is pre-selected for the fit, within a specific domain, by Eqs. (7) and (8). All the experimental points are well arranged along their appropriate curves, i.e. they satisfy Eq. (10) with a very good approximation.

**FIG. 3**. Number of coincidences of the $L = 0$ hadrons $H$ satisfying the rule, Eq. (10), as the trial reference mass varies. Sharp peaks are found exactly at the masses of reference hadrons $h_0$ (marked with ∆), namely: the proton (a) for the ground-state baryons $H$, the pion (b) and rho (c) for the ground-state mesons $H$ plus their radial excitations. The required accuracy for the coincidences is fixed by two threshold values, $T_1$ and $T_2$, which are the permitted maximum distances, respectively, between the mean experimental and calculated mass values of $H$, and between the relative uncertainty intervals. The cumulative count of the coincidences for all these $L = 0$ baryons and mesons, with the reciprocal reinforcement of the peak, is shown in (d), where the results of (a)-(c) are summed together by making the masses of the proton, pion, and rho coincide with the zero of the axis.

**FIG. 4**. Graphs illustrating the networks of the correspondences $H \Rightarrow \{P_i\}$ established by the rule for the ground-state $J^{PC} = 0^{-+}$, $1^{--}$ mesons $H$ (a), and $J^P = \frac{1}{2}^+$ baryons $H$ (b). The trees read top-down: for each hadron $H$, represented by a vertex, the corresponding particles $P_i$ are represented by the adjacent lower vertex/root and their joining edge. A small set of adjunctive, essentially topological, constraints (given in the text) can generate the graphs from their roots upwards, thus correctly reproducing the mass spectrum of these $L = 0$ hadrons $H$. In (a) the edges drawn in bold lead to the mesons $H$ with the lesser $J$. The parentheses indicate hadrons whose masses are unknown.

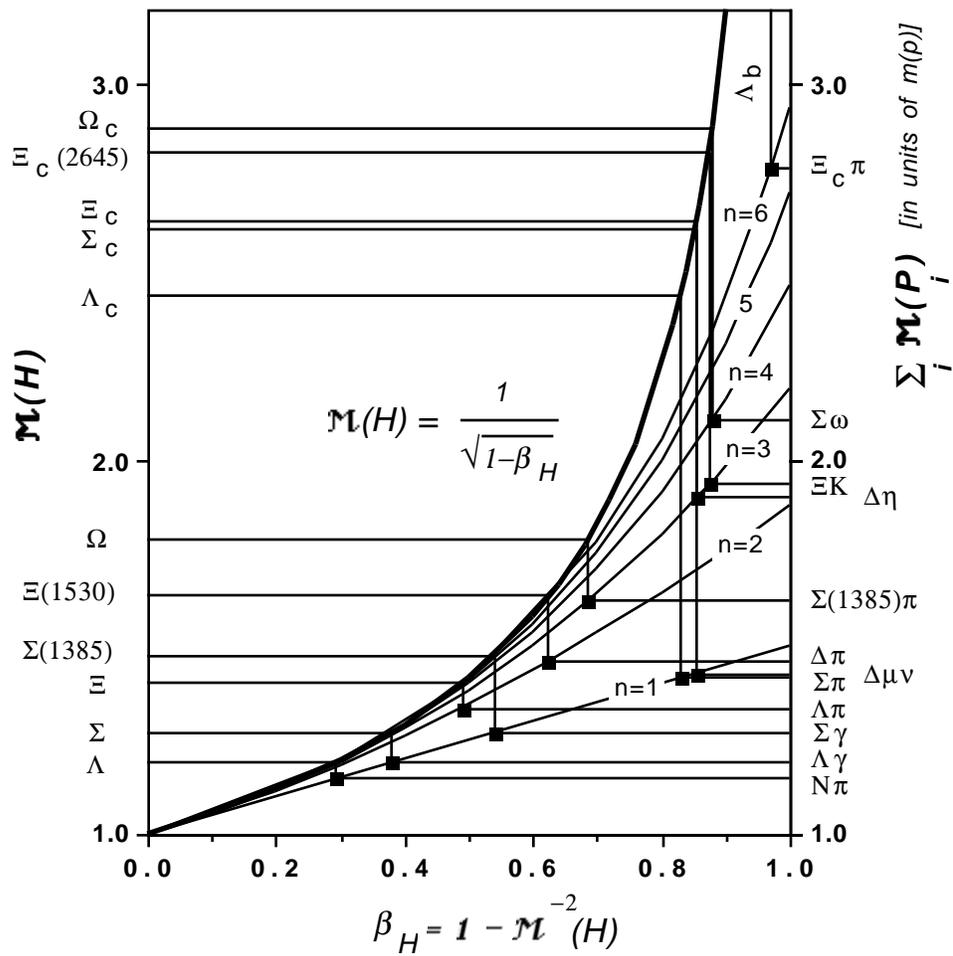

**FIG. 1**

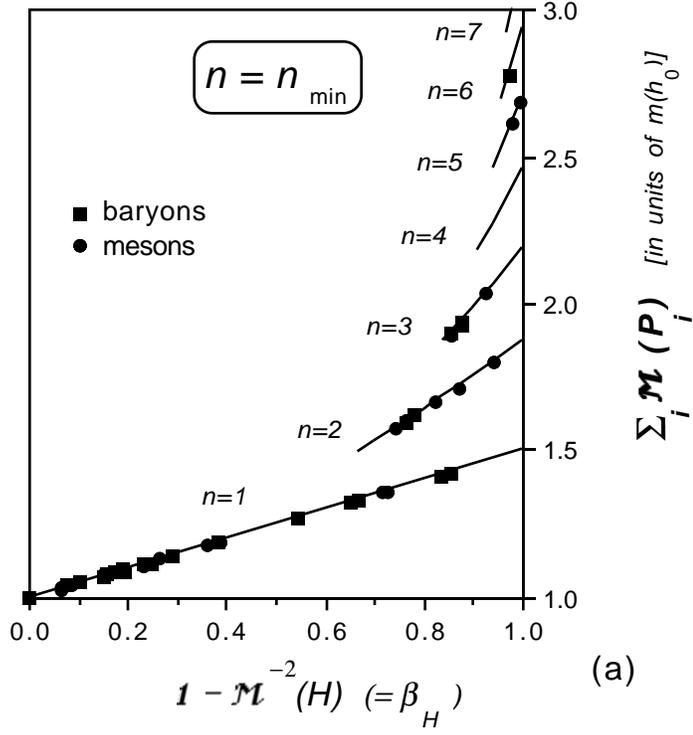

(a)

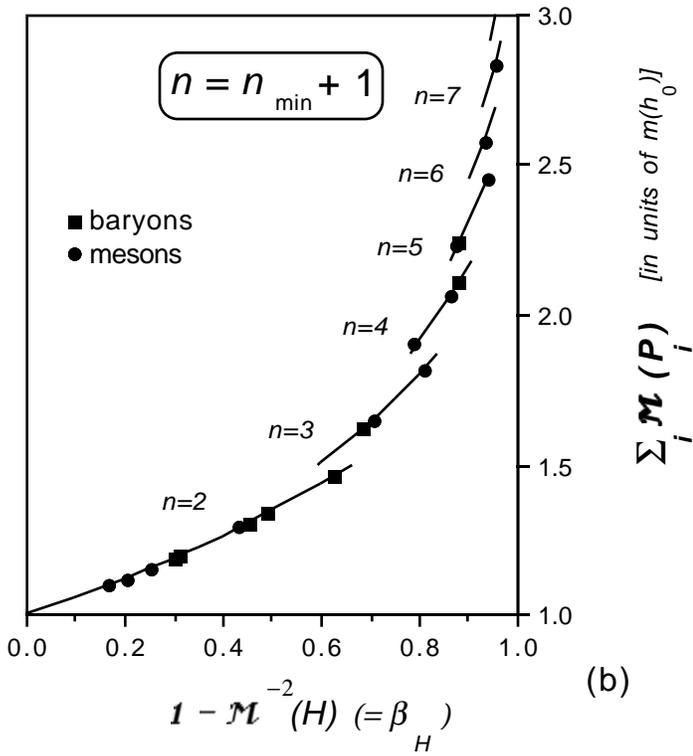

(b)

**FIG. 2**

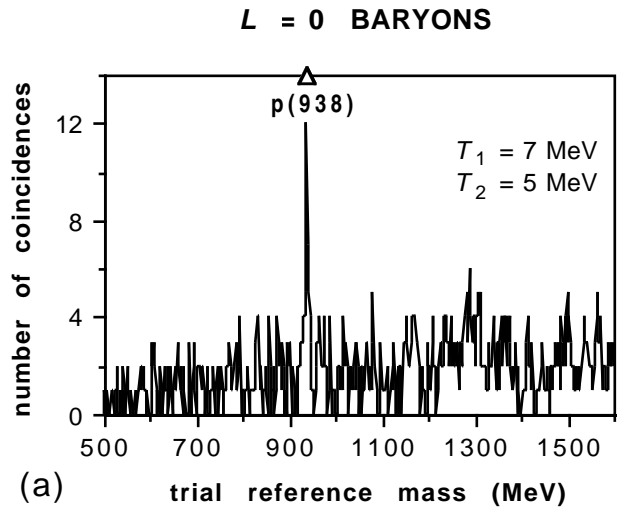

**FIG. 3a**

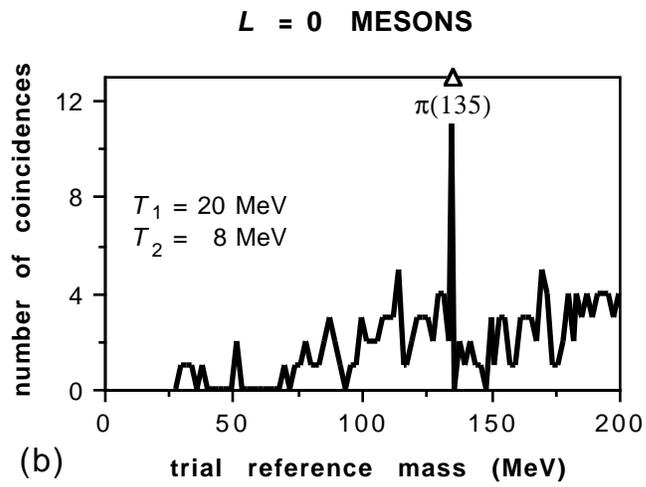

**FIG. 3b**

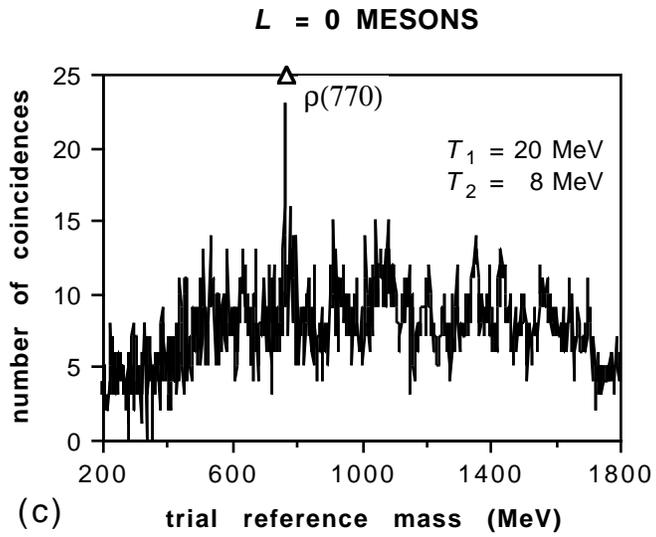

**FIG. 3c**

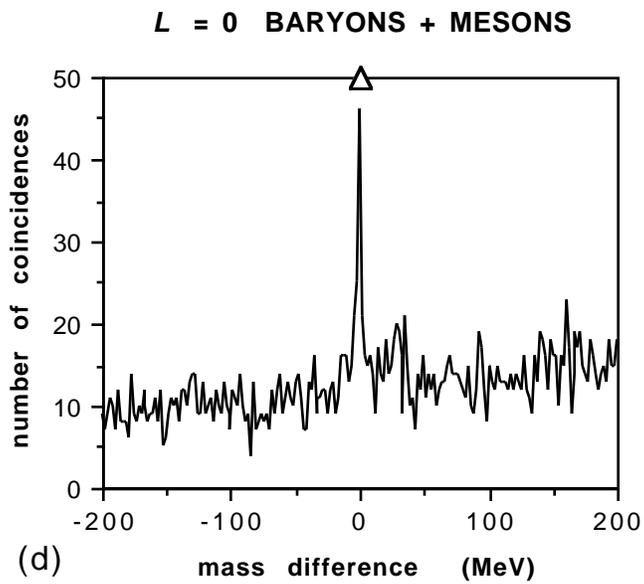

**FIG. 3d**

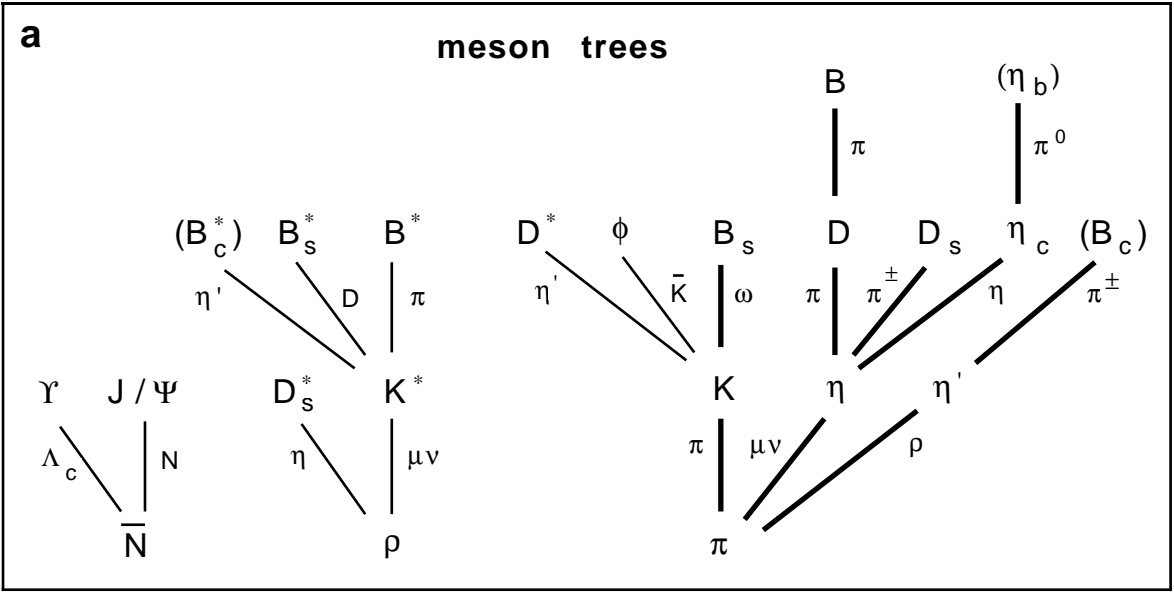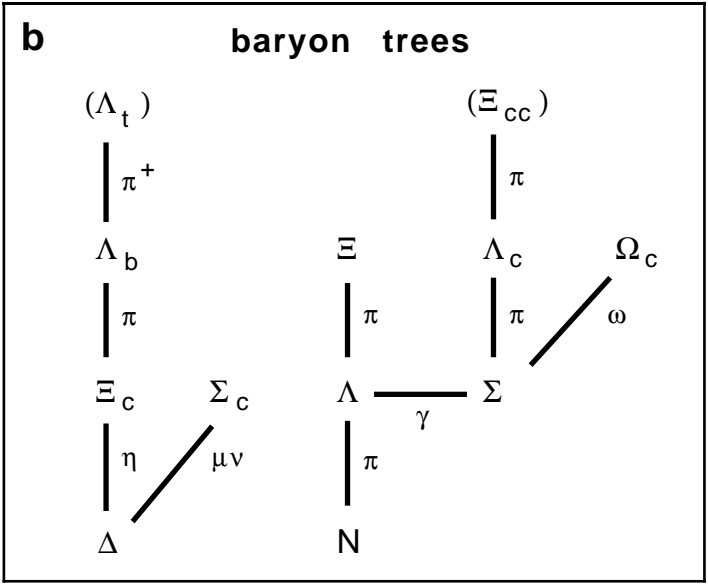

FIG. 4

| $H$ | quark content | $\{P_i\}$ | $h_0$ | $\beta_{H,\text{calc}}$ | $n_{\min}$ | $n$ | $m_{\text{calc}}(H)$ (MeV) | $m_{\text{expt}}(H)$ (MeV) |
|---|---|---|---|---|---|---|---|---|
| $\Lambda$ | $sqq$ | $N\,\pi$ | $p(938)$ | .2940 | 1 | 1 | 1117 | 1116 |
| $\Sigma^0$ | $sqq$ | $\Lambda\,(\gamma)$ | " | .3782 | 1 | 1 | 1190 | 1193 |
| $\Xi$ | $ssq$ | $\Lambda\,\pi$ | " | .4904 | 1 | 2 | 1314 | 1314-1321 |
| $\Sigma(1385)$ | $sqq$ | $\Sigma\,(\gamma)$ | " | .5432 | 1 | 1 | 1388 | 1382-1388 |
| $\Xi(1530)$ | $ssq$ | $\Delta(1232)\,\pi$ | " | .6265 | 1 | 2 | $1535\pm2$ | 1531-1536 |
| $\Omega$ | $sss$ | $\Sigma(1385)\,\pi$ | " | .6877 | 2 | 3 | $1679\pm2$ | 1672 |
| $\Lambda_c$ | $cqq$ | $\Sigma\,\pi$ | " | .83123 | 1 | 1 | $2284\pm1$ | 2285 |
| $\Sigma_c(2455)$ | $cqq$ | $\Delta(1232)\,\mu\,(\nu/e)$ | " | .85400 | 1 | 1 | $2456\pm18$ | 2451-2454 |
| $\Xi_c$ | $csq$ | $\Delta(1232)\,\eta$ | " | .85562 | 3 | 3 | $2469\pm5\;(2467\pm6)$ | 2464-2472 |
| $\Xi_c(2645)$ | $csq$ | $\Xi\,K$ | " | .87455 | 3 | 3 | $2649\pm2$ | $2644\pm2$ |
| or | " | $\Omega\,\pi^+$ | " | .87354 | 3 | 3 | $2638\pm2$ | " |
| $\Omega_c$ | $css$ | $\Xi^0\,\omega(782)$ | " | .87928 | 4 | 5 | $2700\pm3$ | $2704\pm4$ |
| or | " | $\Sigma^0\,\omega(782)$ | " | .87888 | 3 | 4 | $2696\pm1$ | " |
| $\Lambda_b$ | $bqq$ | $\Xi_c\,\pi$ | " | .97228 | 6 | 6 | $5635\pm33$ | $5641\pm50$ |
| $\Omega$ | $sss$ | $\Lambda\,K^-$ | $\Delta(1232)$ | .4527 | 1 | 2 | 1669 | 1672 |
| $\Sigma_c(2530)$ | $cqq$ | $\Sigma\,\omega(782)$ | " | .7610 | 2 | 2 | 2526 | $2530\pm7$ |
| $\Xi_c(2645)$ | $csq$ | $\Delta(1232)\,\rho(770)$ | " | .7816 | 2 | 2 | $2642\pm5$ | $2644\pm2$ |
| $\Lambda(1600)$ | $sqq$ | $N(1440)\,\pi$ | $N(1440)$ | .1907 | 1 | 1 | 1601 | ~1600 |
| $\Sigma(1660)$ | $sqq$ | $\Lambda\,K$ | " | .2379 | 1 | 1 | 1650 | ~1660 |
| $K$ | $s\bar{q}$ | $\pi\,\pi$ | $\pi(135)^0$ | .92667 | 3 | 3 | 498 | 494-498 |
| $\eta$ | $(s\bar{s},q\bar{q})$ | $\pi\,\mu(\nu/e)$ | " | .94029 | 2 | 2 | 552 | 547 |
| $\eta(2S)$ | $(s\bar{s},q\bar{q})$ | $K\,\pi$ | " | .98914 | 18 | 19 | 1295 | $1295\pm4$ |
| $D$ | $c\bar{q}$ | $\eta\,\pi$ | " | .994789 | 21 | 21 | 1870 ($1852\pm7$) | 1864-1870 |
| $D_s$ | $c\bar{s}$ | $\eta\,\pi^\pm$ | " | .995279 | 21 | 21 | 1964 ($1944\pm8$) | 1969 |
| $\eta_c$ | $c\bar{c}$ | $\eta\,\eta$ | " | .997951 | 54 | 55 | 2982 ($2947\pm14$) | $2980\pm2$ |
| [$J/\psi(3097)$ | $c\bar{c}$ | $K\,\bar{K}$ | " | .998114 | 44 | 44[a] | $3108\pm3$ | 3097 ] |
| $\eta_c(2S)$ | $c\bar{c}$ | $\phi(1020)\,\pi^0$ | " | .998588 | 59 | 60 | 3592 | $3594\pm5$ |
| $B^*$ | $b\bar{q}$ | $K^*(892)\,\pi$ | " | .9993521 | 46 | 46 | $5303\pm68$ | $5325\pm2$ |
| $B_s$ | $b\bar{s}$ | $K^0\,\omega(782)$ | " | .99936765 | 71 | 72 | $5368\pm21$ | $5369\pm2$ |
| $K^*(892)$ | $s\bar{q}$ | $\rho(770)\,\mu\,(\nu/e)$ | $\rho(770)$ | .2783 | 1 | 1 | 902 | 891 - 896 |
| $\eta'(958)$ | $(s\bar{s},\,q\bar{q})$ | $\rho(770)\,\pi$ | " | .3601 | 1 | 1 | 958 | 958 |
| $\phi(1020)$ | $s\bar{s}$ | $K\,\bar{K}$ | " | .4406 | 1 | 2 | 1025 | 1019 |
| $\eta(1440)$ | $(s\bar{s},\,q\bar{q})$ | $K\,\rho(770)$ | " | .7044 | 2 | 3 | 1410 | $1415 \pm 10$ |
| $K(2S)$ | $s\bar{q}$ | $K\,\eta$ | " | .7225 | 1 | 1 | 1455   (1452) | ~1460 |
| $\phi(2S)$ | $s\bar{s}$ | $K^0\,\eta'(958)$ | " | .7940 | 3 | 4 | 1689 | $1680 \pm 20$ |
| $\eta(3S)$ | $(s\bar{s},\,q\bar{q})$ | $K\,\bar{K}^*(892)$ | " | .8073 | 2 | 3 | $1746 \pm 1$ | $1760 \pm 11$ |
| $K(3S)$ | $s\bar{q}$ | $K\,\omega(782)$ | " | .8239 | 2 | 2 | $1827 \pm 1$ | ~1830 |
| $D^*(2010)$ | $c\bar{q}$ | $K\,\eta'(958)$ | " | .8544 | 3 | 3 | $2009 \pm 1$ | 2006-2011 |
| $D_s^*(2110)$ | $c\bar{s}$ | $\rho(770)^{\pm}\,\eta$ | " | .8681 | 2 | 2 | 2111   (2107) | 2112 |
| [$\eta_c$ | $c\bar{c}$ | $\phi(1020)\,\eta'(958)$ | " | .93348 | 5 | 6 | $2972 \pm 1$ | $2980 \pm 2$ ] |
| $J/\psi(3097)$ | $c\bar{c}$ | $N\,\bar{N}$ | " | .93857 | 4 | 5 | 3093 | 3097 |
| [$\eta_c(2S)$ | $c\bar{c}$ | $N\,\bar{\Delta}(1232)$ | " | .95430 | 6 | 7 | $3586 \pm 5$ | $3594 \pm 5$ ] |
| $\psi(2S)$ | $c\bar{c}$ | $\Sigma\,\bar{\Sigma}$ | " | .95695 | 8 | 9 | $3695 \pm 1$ | 3686 |
| $B$ | $b\bar{q}$ | $D\,\pi$ | " | .978905 | 5 | 5 | $5278 \pm 20$ | 5277-5281 |
| $B_s^*$ | $b\bar{s}$ | $D\,K^*(892)$ | " | .980011 | 10 | 11 | $5422 \pm 11$ | $5416 \pm 3$ |
| $\Upsilon(9460)$ | $b\bar{b}$ | $p\,\bar{\Lambda}_c$ | " | .993424 | 14 | 14 [a] | $9454 \pm 29$ | 9460 |
| or | " | $\Sigma(1385)^0\,\bar{\Omega}_c$ | " | .993438 | 23 | 24 | $9464 \pm 115$ | " |
| $\Upsilon(2S)$ | $b\bar{b}$ | $n\,\bar{\Lambda}$ | " | .994157 | 5 | 5 [a] | $10029 \pm 2$ | 10023 |
| $\Upsilon(3S)$ | $b\bar{b}$ | $D^*(2010)^{\pm}\,\bar{D}_s^*(2110)$ | " | .9945328 | 23 | 24 | $10368 \pm 36$ | 10355 |
| $\Upsilon(4S)$ | $b\bar{b}$ | $\Omega\,\bar{\Omega}$ | " | .9947349 | 15 | 15 [a] | $10565 \pm 36$ | $10580 \pm 4$ |

**TABLE I**

| $H$ | $L$ | quark content | $\{P_i\}$ | $h_0$ | $\beta_{H,\text{calc}}$ | $n_{\min}$ | $n$ | $m_{\text{calc}}(H)$ (MeV) | $m_{\text{expt}}(H)$ (MeV) |
|---|---|---|---|---|---|---|---|---|---|
| $\Lambda(1520)$ | 1 | $sqq$ | $\Sigma(1385)\,\pi$ | $N(1520)$ | .0024 | 1 | 1 | 1522 | 1520 |
| $\Sigma(1670)$ | 1 | $sqq$ | $\Lambda(1520)\,\pi$ | " | .1800 | 1 | 1 | $1679 \pm 2$ | 1665-1685 |
| $\Lambda(1690)$ | 1 | $sqq$ | $\Lambda\,\eta$ | " | .1974 | 1 | 1 | 1687 | 1685-1695 |
| $\Xi(1820)$ | 1 | $ssq$ | $\Sigma(1670)\,\pi$ | " | .3072 | 1 | 2 | $1826 \pm 12$ | $1823 \pm 5$ |
| $\Lambda_c(2625)$ | 1 | $cqq$ | $\Xi(1530)^0 K^+$ | " | .6651 | 1 | 1 | $2627 \pm 1$ | 2626 |
| $\Lambda(1405)$ | 1 | $sqq$ | $\Sigma(1385)^0 (\gamma)$ | $N(1535)$ | -.1971 | 1 | 1 | $1403 \pm 1$ | $1407 \pm 4$ |
| $\Sigma(1620)$ | 1 | $sqq$ | $\Lambda\,K$ | " | .0995 | 1 | 1 | 1618 | ~1620 |
| $\Lambda(1670)$ | 1 | $sqq$ | $N(1520)\,\pi$ | " | .1593 | 1 | 1 | $1674 \pm 13$ | 1660-1680 |
| $\Lambda_c(2593)$ | 1 | $cqq$ | $N(1535)\,K$ | " | .6458 | 1 | 1 | 2579 [b] | 2594 |
| $\Lambda(1820)$ | 2 | $sqq$ | $\Sigma(1670)\,\pi$ | $N(1680)$ | .1592 | 1 | 1 | $1824 \pm 13$ | 1815-1825 |
| $\Sigma(1915)$ | 2 | $sqq$ | $\Sigma(1385)\,K$ | " | .2384 | 1 | 1 | 1925 | 1900-1935 |
| $\Xi(2030)$ | 2 | $ssq$ | $\Delta(1232)\,\rho(770)$ | " | .3099 | 1 | 2 | $2022 \pm 3$ | $2025 \pm 5$ |
| $\Sigma(2030)$ | 2 | $sqq$ | $\Xi(1530)\,K$ | $\Delta(1950)$ | .0811 | 1 | 1 | $2034 \pm 1$ | 2025-2040 |
| $f_0(980)$ | 1 | $(s\bar{s}, q\bar{q})$ | $\omega(782)\,\pi^0$ | $\rho(770)$ | .3853 | 1 | 1 | 981 | $980 \pm 10$ |
| or | " | " | $\rho(770)\,\mu\,(\nu/e)$ | $b_1(1235)$ | -.5804 | 1 | 1 | $979 \pm 1$ | " |
| $h_1(1170)$ | 1 | $(s\bar{s}, q\bar{q})$ | $\phi(1020)\,\pi^0$ | " | -.1243 | 1 | 1 | 1161 | $1170 \pm 20$ |
| $K_1(1270)$ | 1 | $s\bar{q}$ | $K\,\rho(770)$ | " | .0529 | 1 | 1 | $1265 \pm 1$ | $1273 \pm 7$ |
| or | 1 | " | $K\,\omega(782)$ | " | .0759 | 1 | 1 | 1280 | " |
| $h_1(1380)$ | 1 | $(s\bar{s}, q\bar{q})$ | $b_1(1235)\,\pi$ | " | .1946 | 1 | 2 | 1372 | $1380 \pm 20$ |
| $K_1(1400)$ | 1 | $s\bar{q}$ | $b_1(1235)\,\pi$ | " | .2230 | 1 | 1 | 1396 | $1402 \pm 7$ |
| $D_1(2420)^0$ | 1 | $c\bar{q}$ | $a_0(980)\,\eta'(958)$ | " | .7417 | 2 | 2 | $2422 \pm 4$ | $2422 \pm 2$ |
| $D_{s1}(2536)$ | 1 | $c\bar{s}$ | $D_s\,(\gamma)$ | " | .76284 | 2 | 2 | $2528 \pm 6$ | 2535 |
| $h_c(1P)$ | 1 | $c\bar{c}$ | $a_2(1320)^0 f_1(1420)$ | " | .87785 | 4 | 5 | $3522 \pm 11$ | 3526 |
| $K_2^*(1430)$ | 1 | $s\bar{q}$ | $f_2(1270)\,\pi$ | $a_2(1320)$ | .1426 | 1 | 1 | $1423 \pm 2$ | 1424-1434 |
| or | " | " | $f_1(1285)\,\pi$ | " | .1538 | 1 | 1 | $1433 \pm 1$ | " |
| $f_2'(1525)$ | 1 | $s\bar{s}$ | $\phi(1020)\,K^0$ | " | .2537 | 1 | 2 | 1526 | $1525 \pm 5$ |
| $D_2^*(2460)$ | 1 | $c\bar{q}$ | $K^*(892)\bar{K}^*(892)$ | " | .7125 | 1 | 1 | $2458 \pm 4$ | 2455-2463 |
| $\chi_{c2}(1P)$ | 1 | $c\bar{c}$ | $a_2(1320)\,K_1(1400)$ | " | .86313 | 3 | 4 | $3563 \pm 27$ | 3556 |
| $\chi_{b2}(1P)$ | 1 | $b\bar{b}$ | $f_2(1270)\,\chi_{c1}(1P)$ | " | .9823286 | 10 | 11 | $9915 \pm 23$ | 9913 |
| $K_3^*(1780)$ | 2 | $s\bar{q}$ | $a_0(980)\,\omega(782)$ | $\rho_3(1690)$ | .0880 | 1 | 1 | $1771 \pm 1$ | $1770 \pm 10$ |
| $\phi_3(1850)$ | 2 | $s\bar{s}$ | $\eta'(958)\,K^*(892)^0$ | " | .1707 | 1 | 2 | 1857 | $1854 \pm 7$ |

**TABLE II**

| $H$ | quark content | $\{P_i\}$ | $h_0$ | $\beta_{H,\text{calc}}$ | $n_{\text{min}}$ | $n$ | $m_{\text{calc}}(H)$ (MeV) |
|---|---|---|---|---|---|---|---|
| $B_c$ | $b\bar{c}$ | $\eta'(958)\,\pi^{\pm}$ | $\pi(135)^0$ | .9995356 | 52 | 52 | $6263 \pm 50$ |
| $B_c^*$ | $b\bar{c}$ | $K^*(892)^{\pm}\eta'(958)$ | $\rho(770)$ | .984973 | 4 | 4 | $6254 \pm 33$ |
| $\eta_b$ | $b\bar{b}$ | $\eta_c\,\pi^0$ | " | .9931272 | 13 | 13[a] | $9248 \pm 108$ |
| $\Xi_{cc}$ | $ccq$ | $\Lambda_c\,\pi$ | $p(938)$ | .9340 | 5 | 6 | $3652 \pm 4$ |

**TABLE III**